\documentclass[spthm,floatrow]{iacrtrans}
\usepackage[utf8]{inputenc}
\usepackage{multirow}
\usepackage{algorithm}  
\usepackage{algpseudocode}  
\usepackage{amsmath}  
\author{Yuanchao Ding\inst{1,2},Hua Guo\inst{1,2,*},Yewei Guan\inst{1,3}, Hutao Song\inst{1,3},Xiyong Zhang\inst{4} }
\institute{
  School of Cyber Science and Technology, Beihang University, Beijing 100191, China
  \and
   State Key Laboratory of Cryptology, P.O. Box 5159, Beijing 100878, China
   \and
   State Key Laboratory of Software Development Environment, Beihang University, Beijing 100191, China
   \and
   Beijing Institute of Satellite Information Engineering, Beijing 100086, China
}
\title{Some New Methods to Generate Short Addition Chains}
\begin{document}

\maketitle

\let\thefootnote\relax
\footnotetext{*Corresponding Author}
\footnote{Email addresses:}
\footnote{dych21@buaa.edu.cn (Yuanchao Ding)}
\footnote{hguo@buaa.edu.cn (Hua Guo)}
\footnote{ame\_reiori@buaa.edu.cn (Yewei Guan)}
\footnote{htsong@buaa.edu.cn (Hutao Song)}
\footnote{Xiyong.Zhang@hotmail.com (Xiyong Zhang)}

%%%% 5. KEYWORDS %%%%

%%%% 6. ABSTRACT %%%%
\begin{abstract}
Modular exponentiation and scalar multiplication are important operations of most public key cryptosystems, and their fast calculation is essential to improve the system efficiency. The shortest addition chain is one of the most important mathematical concepts to realize the optimization. However, finding a shortest addition chain of length \emph{k} is generally regarded as an \textbf{NP}-hard problem, whose time complexity is comparable to $O(k!)$. This paper proposes some novel methods to generate short addition chains. We firstly present a Simplified Power-tree method by deeply deleting the power-tree whose time complexity is reduced to $O(k^2)$. Moreover, a Cross Window method and its variant are introduced by improving the Window method. More precisely, the Cross Window method uses the cross correlation to deal with the windows and its pre-computation is optimized by a new Addition Sequence algorithm. The theoretical analysis is conducted to show the correctness and effectiveness. Meanwhile, the experiment shows that the new methods can obtain shorter addition chains compared to the existing methods. The Cross Window method with the Addition Sequence algorithm can attain 44.74\% and 9.51\% reduction of the addition chain length, in the best case, compared to the Binary method and the Window method respectively.

\end{abstract}

%%%% 7. PAPER CONTENT %%%%
\section{Introduction\label{chap:1}}
Public-key cryptosystem is widely used, but the speed is much lower than symmetric cryptosystem. In the process of encryption and decryption, modular exponentiation and scalar multiplication are the key factors affecting the efficiency. Common public key cryptosystems include DH, RSA, ElGamal, ECC, etc. Modular exponentiation is used in DH, RSA and ElGamal, which is
\begin{equation}
y=x^e mod \,c.
\end{equation}
In ECC, scalar multiplication is used as
\begin{equation}
Q=eP.
\end{equation}

In these two operations, the calculation of positive integer \textit{e} is involved. In fact, the operations can be abstractly approached as an Addition Chain Problem (ACP) to find a shortest Addition Chain (AC). There are different calculation paths to get \textit{e}. For example, when $e=9$, there are 1-2-4-8-9, 1-2-3-6-9, \dots, etc. Such paths are called the addition chains of \textit{e}. 
\par
Given a positive integer \textit{e}, an addition chain \textit{A} of $e$ with length \textit{r} is a sequence of positive integers: $A=a_0, a_1, a_2, \dots, a_r$, where $a_0=1, a_1=2,\dots, a_r=e$, and for all $k \ge1, a_k=a_i+a_j, k > i \ge j$. When $i=j=k-1$, this step is called doubling step (i.e. $a_k=2a_{k-1}$). When $i=k-1$, this step is called star step (i.e. $a_k=a_{k-1}+a_j, j<k$). In this paper, $n(e)=\lfloor \log_2{e}\rfloor+1$ indicates $e$ is an $n$-bit integer. $h(e)$ is the Hamming weight of \textit{e} which means the number of 1s in the binary form of $e$. $l(e)$ is the shortest addition chains length of \textit{e}.
\par

In the practical public key cryptosystem, the operands are usually selected with long bits for security. For example, RSA uses 1024, 2048 or 4096 bit, which costs a long time for the computer to perform the operation. A shorter addition chain means faster
execution of the corresponding modular exponentiation or scalar multiplication, because there is one-to-one
relationship between an element of the addition chain and the computation of modular exponentiation or scalar multiplication. For example, when $e=9$, the addition chain 1-2-4-8-9 corresponds to $x^9=((x^2)^2)^2x=x\mbox{-}x^2\mbox{-}x^4\mbox{-}x^8\mbox{-}x^9$ or $9P=2(2(2P))+P=P\mbox{-}2P\mbox{-}4P\mbox{-}8P\mbox{-}9P$. However, ACP is generally regarded as an \textbf{NP}-hard problem\cite{kunihiro2000new,DBLP:conf/cis/CortesRJC05,DBLP:journals/jcsci/NomaMMZ17}. Therefore, optimizing the result of addition chain to improve the execution speed of modular exponentiation or scalar multiplication is of great practical significance to improve the efficiency of public key cryptosystems.
\par
At present, the methods of generating addition chain can be classified into two main types. The first type \cite{DBLP:journals/cluster/BahigA18,DBLP:journals/computers/BahigK19,DBLP:journals/dm/ThurberC21}, called minimal length addition chain(MLAC), focuses on generating optimal addition chain, while the second one, called short addition chain(SAC), aims to generate short addition chain. The length of generated chain in the first type is minimal, which can well optimize the modular exponentiation and scalar multiplication. However, as mentioned above, the running time increases rapidly. In the second type, we can strike a balance between the length of generated chain and performance, which is more practical. In this work, we will focus on the second type of methods. 
\par

For the second type, many methods have been proposed, such as the Binary method, the \textit{m}-ary ($2^k$-ary) method \cite{brauer1939addition}, the Window method \cite{knuth2014art}, the Power-tree method \cite{knuth2014art}, the Genetic algorithm \cite{DBLP:conf/cis/CortesRJC05,DBLP:conf/cec/Vazquez-Fernandez16}, the Artificial Immune System \cite{DBLP:journals/tec/CortesRC08}, and the Evolutionary Programming \cite{DBLP:journals/eaai/Dominguez-IsidroMO15,DBLP:journals/heuristics/PicekCJM18}, etc. The Binary method is widely used in fast modular exponentiation and scalar multiplication, which can be further optimized. The \textit{m}-ary method\cite{brauer1939addition} divides the binary form of integers into windows, which are included in a pre-computation. Then it proceeds by scanning all the windows. The Window method\cite{knuth2014art} can achieve better results than \textit{m}-ary method by looking for the windows whose head and tail are non-zero, thus to reduce the pre-computation. The Power-tree method\cite{knuth2014art}, the Genetic Algorithm\cite{DBLP:conf/cis/CortesRJC05,DBLP:conf/cec/Vazquez-Fernandez16}, the Artificial Immune System\cite{DBLP:journals/tec/CortesRC08} and the Evolutionary Programming\cite{DBLP:journals/eaai/Dominguez-IsidroMO15,DBLP:journals/heuristics/PicekCJM18} need complex operations, which are suitable for relatively small integers until now. The Window-based methods are feasible to solve large integers within a short time and have been repeatedly optimized.
\par 
Scholars have made a number of analysis and improvements on the Window method \cite{DBLP:conf/crypto/BosC89,kunihiro1998window,kunihiro2000new,DBLP:journals/ecs/Abd-EldayemAF10,DBLP:conf/cisc/KozielAJK16}, which are collectively called Window-based methods. In 1989, Bos and Coster\cite{DBLP:conf/crypto/BosC89} filled the pre-computation with addition sequences instead of all odd numbers, supporting bigger window size and smaller length of result. Kunihiro and Yamamoto\cite{kunihiro1998window} further optimized the size of the pre-computation using Tunstall code. In 2000, Kunihiro and Yamamoto\cite{kunihiro2000new} proposed the run-length method, which worked well when processing integers with large hamming weight. They also proposed the Hybrid method, a hybrid of Run-length method and Window method. In 2010, Mohamed et al.\cite{DBLP:journals/ecs/Abd-EldayemAF10} proposed an algorithm based on Window method with small width by using 2's Complements for finding addition-subtraction chain of 160-bit integers. In 2016, Brian et al. \cite{DBLP:conf/cisc/KozielAJK16} used Window method to find addition chain of smooth isogeny primes.
\par
In this paper, we take into account some interesting     observations about existing methods and put forward our novel ideas. To obtain a shortest addition chain, the exhaustive search of the Power-tree method is unpractical. We study the Binary method carefully and find a ``backward-adding'' way to construct the addition chain with our simplified power-tree, which deletes the nodes in power-tree substantially. For the Window method, adjacent windows are usually adopted which has limited window combinations. We exploit cross windows which diversify the window combinations and present our Cross Window method for better results. In the pre-computation, a new Addition Sequence algorithm is presented to build a shorter pre-computation of the used windows, which leads to our Cross Window method with the Addition Sequence algorithm.
\par
To sum up, the contributions of this paper are as follows:
\par
1. Firstly, we present a Simplified Power-tree method by deeply deleting the power-tree,
whose time complexity is reduced from $O(k!)$ to $O(k^2)$ .
\par
2. Secondly, a Cross Window method is introduced by
improving the Window method, which achieves better result since more  window combinations are handled by using cross windows. 

3. Thirdly, a Cross Window method with the Addition Sequence algorithm is given to optimize the pre-computation in the Cross Window method with using our new Addition Sequence algorithm, which attains 9.51\% reduction of the addition chain length, in the best case, compared to the Window method.

\par
This paper is structured as follows.
 In Section \ref{chap:1}, we present some basic introduction of the state of the art as well the main contributions
of this paper. Section \ref{chap:2}
briefly reviews some general existing methods. Section \ref{chap:3} shows a detailed description of our
novel methods, including the Simplified Power-tree method, the Cross Window method and the Cross Window method with Addition Sequence algorithm.
We perform our experiment in Section \ref{chap:4}, which shows the new methods can obtain shorter addition chains compared to the existing methods. Section \ref{chap:5} concludes the whole paper.

\section{Existing methods\label{chap:2}}

\subsection{Binary Method}
The Binary method (BM) uses binary representation of an integer, and an optional addition is performed depends on whether a bit is 1 or 0.
The general implementation of BM is shown in Alg. \ref{alg1} (in \cite{DBLP:journals/jcsci/NomaMMZ17}).
\par
    \begin{algorithm}[htb]
        \caption{Binary Method}
        \begin{algorithmic}[1] %每行显示行号
        \Require $e=(e_{n-1}\dots e_1e_0)_2$
        \Ensure $A = \{a_0, a_1, a_2, \dots, a_r=e\}$
        \State $A = \{a_0=1\}, r=1$
        \For {$i$ from $n-2$ down to 0}
        \State $A = A \cup \{ a_{r}:= 2a_{r-1}\}$, $r=r+1$
        \If{$e_i=1$}
        \State $A = A \cup \{ a_{r}:= a_{r-1} +1 \}$, $r=r+1$
        \EndIf
        \EndFor
        \State \Return $A$
        \end{algorithmic}
        
        \label{alg1}
    \end{algorithm}
\par

\subsection{Power-tree Method}
\begin{figure}[H]
\centering
\ffigbox{\includegraphics[width=0.95\textwidth]{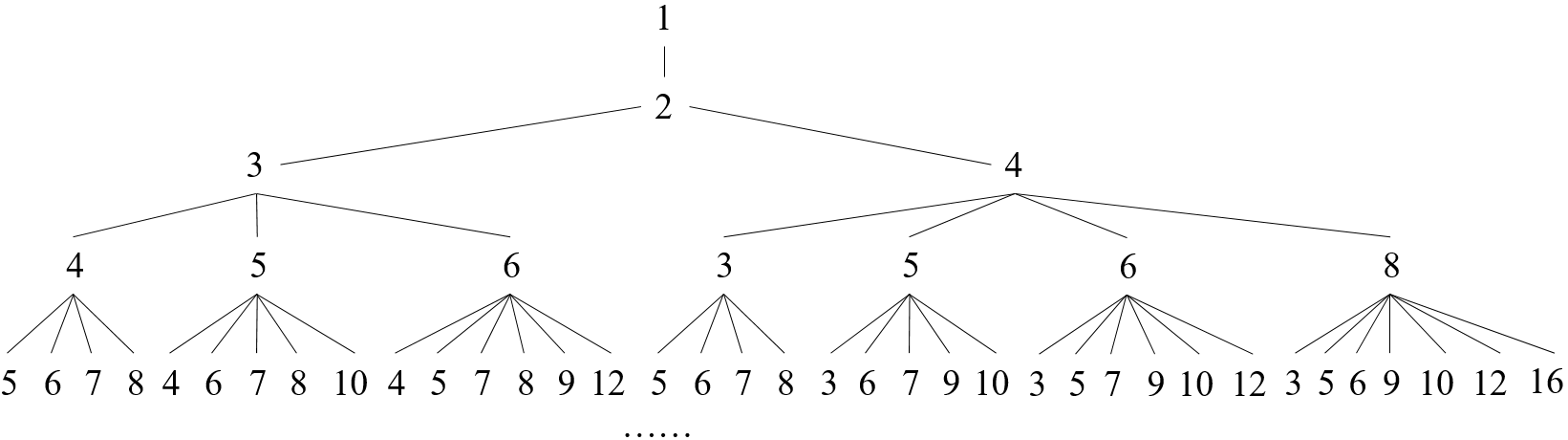}}
{\caption{Power-tree.}
\label{fig1}}
\end{figure}

The Power-tree method (PTM) means that all nodes are represented in the form of a tree, and the nodes on the path are used as the addition chain of an integer. A complete power-tree without duplicate nodes on any path is a tree that contains all possible results, as shown in Fig. \ref{fig1}.

\par

The shortest addition chain of an integer can be determined by exhaustive search within all paths, which takes a long time. Let the depth of node 1 be 0, the number of subnodes nodes in depth \textit{k} is not less than $k+1$, and the total number of nodes in depth $k$ is over $1\cdot 2 \cdot 3 \cdot  \dots \cdot k=k!$.

\subsection{Window Method } 
The idea of the Window method (WM) is to split the 
binary form of an integer into some windows, then the windows are processed to get the addition chain through two parts: pre-computation and construction. Let the window length be \textit{k}. Pre-computation selects all odd integers from 1 to $2^k-1$, and 2, which is \{1, 2, 3, 5, 7, \dots, $2^k-1$\}, with length $2^{k-1}+1$.
\par
In WM, for the binary form of an integer $e$, we read a window $w$ (the bit-length of $w$ (denote as) $n(w)\le k$ and $MSB(w)=LSB(w)=1$, where $MSB(w)$ and $LSB(w)$ indicate the most and the least 
significant bit of $w$ respectively). As a result, \textit{w} is in the pre-computation. In construction, $n(w)$ times doubling step and one time star step are performed. For the consecutive 0s, doubling steps are directly conducted.
The implementation of WM is shown in Alg. \ref{alg2}.

    \begin{algorithm}[htb]
        \caption{Window Method}
        \begin{algorithmic}[1] 
        \Require $k,e=(e_{n-1}\dots e_1e_0)_2$
        \Ensure $A = \{a_0, a_1, a_2, \dots, e\}$
        \State $A = \{1, 2, 3, 5, 7, \dots, 2^k-1\}$
        \State $i = n-1, r=2^{k-1}$
        \State find the longest bitstring $e_ie_{i-1}\dots e_j$ such that $i-j+1\le k$ and $e_j=1$
        \State $t =(e_ie_{i-1}\dots e_j)_2,i=j-1$
        \While {$i \ge 0$}
        \If{$e_i=0$}
        \State $A = A \cup \{ a_{r}:=2t\}, r=r+1, t=a_{r-1}$
        \Else
        \State find the longest bitstring $e_ie_{i-1}\dots e_j$ such that $i-j+1\le k$ and $e_j=1$
        \For{$t$ from $i-j+1$ down to 1}
        \State $A = A \cup \{ a_{r}:=2t\}, r=r+1, t=a_{r-1}$
        \EndFor
        \State $A = A \cup \{ a_{r}:=t+(e_ie_{i-1}\dots e_j)_2\}, r=r+1, t=a_{r-1}$
        \State $i=j-1$
        \EndIf
        \EndWhile
        \State \Return $A$
        
        \end{algorithmic}
        
        \label{alg2}
    \end{algorithm}

\par

\section{New methods\label{chap:3}}
In this section, we propose the Simplified Power-tree Method (SPTM), 
the Cross Window method (CWM) and its variant the Cross Window method with the Addition Sequence algorithm (CWM-ASA). The former improves PTM by deleting the power-tree. The latters improve WM by using cross windows and optimizing the pre-computation.
\subsection{Simplified Power-tree Method }
We first give another view of BM. Instead of using an optional addition mixed in doubling steps, we do the optional addition when all the doubling steps are done and the addition number 1 is 
adjusted to corresponding numbers. This implementation of BM is Alg. \ref{alg3}. This view of BM shows a feasible way to construct the addition chain, which leads to the key point  ``backward-adding'' of our Simplified Power-tree method.
    \begin{algorithm}[htb]
        \caption{Binary Method$^*$}
        \begin{algorithmic}[1] %每行显示行号
        \Require $e=(e_{n-1}\dots e_1e_0)_2$
        \Ensure $A = \{a_0, a_1, a_2, \dots, a_r=e\}$
        \State $A = \{1,2,4,\dots,2^{n-1}\},r = n$
        \For {$i$ from $n-2$ down to 0}
        \If{$e_i=1$}
        \State $A = A \cup \{ a_{r}:= a_{r-1} +a_{i} \}$, $r=r+1$
        \EndIf
        \EndFor
        \State \Return $A$
        \end{algorithmic}
        
        \label{alg3}
    \end{algorithm}

\begin{figure}[htb]
\centering
\ffigbox{\includegraphics[width=0.8\textwidth]{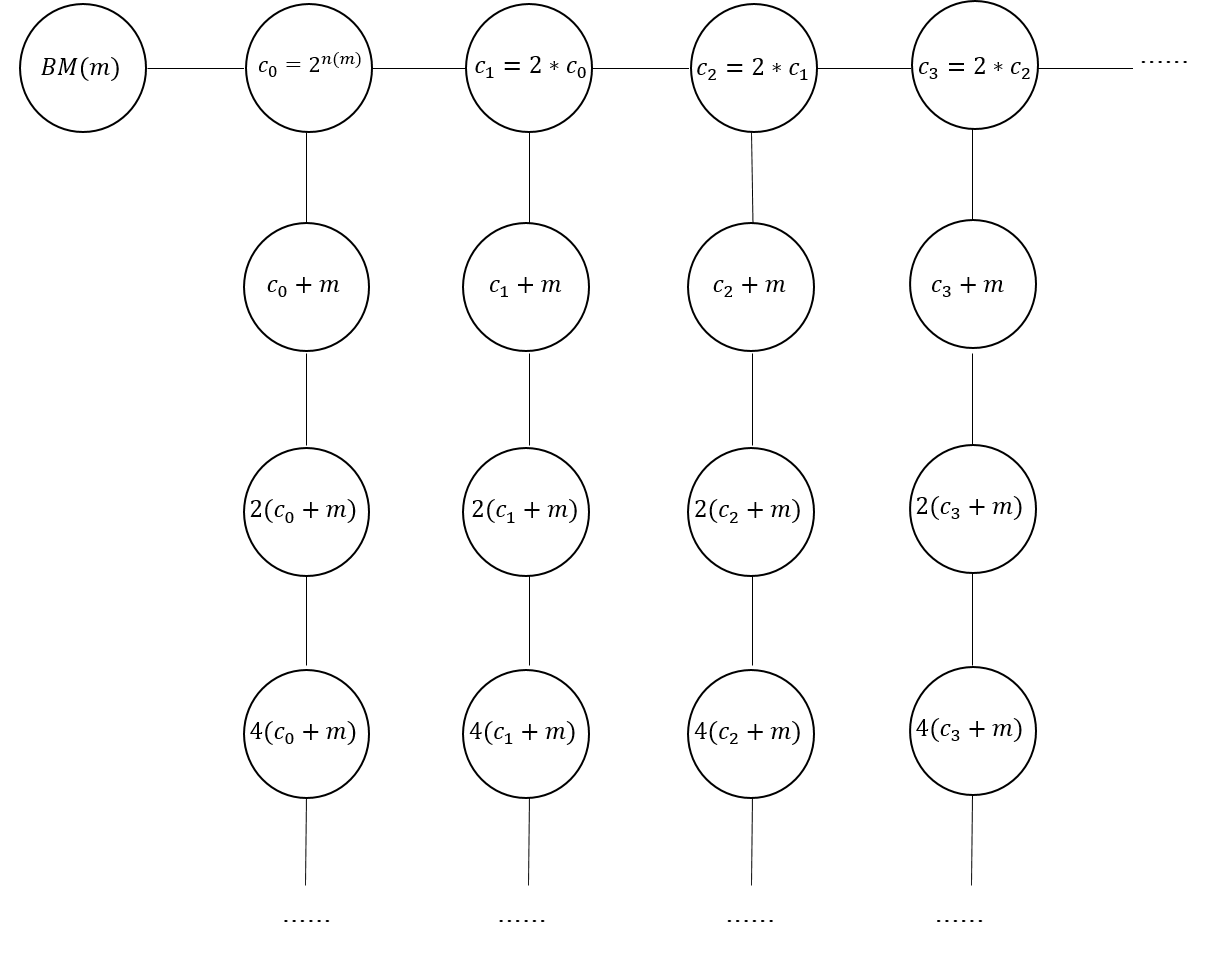}}
{\caption{Simplified power-tree.}
\label{fig2}
}
\end{figure}

SPTM is proposed by subtly deleting tree nodes, which results in relatively small time and space complexity.
A simplified power-tree consists of root chain, main chain and branch chains. The structure of the root chain is BM($m$) where $m$ is a base integer to build branch chains, and the main chain is $\{2^i|n(m)\le i \le t, 2^{t}\le e <2^{t+1}\}$. For each node $2^i$ in the main chain, a branch chain follows as  $\{2^j(2^i+m)|0\le j \le t,2^{t}(2^i+m)\le e < 2^{t+1}(2^i+m)\}$. The structure of simplified power-tree is shown in Fig. \ref{fig2}.

\par

Based on the simplified power-tree, the steps of constructing the addition chain of \textit{e} are as follows:
\par
(1) Obtain an addition chain of $e$ by BM($e$) and record it as the result.
\par
(2) Search the branch chains and update the recorded addition chain whenever get a shorter addition chain.
\par
(3) Output the recorded addition chain.
\par
More specifically in step (2), for the branch chain followed $c_i$, the corresponding addition chain of  \emph{e} is directly obtained if \emph{e} is in the branch chain. Otherwise we form an initial chain as BM($m$) $\cup$ $\{2^j|n(m)\le j \le i\}$ $\cup$ $\{2^j(2^i+m)|0\le j \le t,2^{t}(2^i+m)\le e < 2^{t+1}(2^i+m)\}$  and do the ``backward-adding'': whenever the newest integer in current chain adding the node backward in the initial chain is less than \textit{e}, do the adding and append the adding result in current chain. Each branch chain can get an addition chain of \textit{e}, as proved in Theorem 1. Then we search all chains and update the recorded addition chain whenever we get a shorter addition chain.
The implementation of SPTM is shown in Alg. \ref{alg4}.\par
    \begin{algorithm}[htb]
        \caption{Simplified Power-tree Method}
        \begin{algorithmic}[1] %每行显示行号
        \Require \textit{m,e}
        \Ensure $A = \{a_0, a_1, a_2, \dots, e\}$
        \State $A=BM(e)$
        \For {$i$ from $n(m)$ to $n(e)-1$}
        \State $T=BM(m) \cup \{{2^{n(m)},2^{n(m)+1},\dots,2^{i},2^{i}+m}\},r=length(T)-1$
        \While{$t_r<e$}
        \State $r=r+1,T = T \cup \{ t_{r}:= 2t_{r-1} \}$
        \EndWhile
        \If{$t_r=e$ and $r<length(A)$}
            \State $A=T$
        \Else
				\State $r=r-1$
            \For{$j$ from $r-1$ down to $0$}
                \If{$t_r=e$ and $r<length(A)$}
                \State $A=T$
                \EndIf
                \If{$t_r+t_j<e$}
                \State $r=r+1,T = T \cup \{ t_{r}:= t_{r-1}+t_j \}$

                \EndIf
            \EndFor
        \EndIf
        \EndFor
        \State \Return $A$
        \end{algorithmic}
				\label{alg4}
    \end{algorithm}

\par 
SPTM is exactly efficient. For any given positive integer $e$, because the main chain and branch chains mainly contain doubling steps, their lengths are approximately equal to the bit-length of $e$ (i.e. $O(\log e))$. Therefore, the time complexity of SPTM is $O((\log e)^2)$. The recorded addition chain is constantly  updated so that the space complexity is $O(\log e)$.
\par
\textbf{Theorem 1:}
For any given positive integer, SPTM can produce an addition chain. 
\par
\textbf{Proof: }For the main chain, an addition chain of \textit{e} is generated by BM. For the branch chain followed $c_i$, let $e=t(c_i+m)+C$, where $0 \le C<c_i+m$. In fact, it has $0\le C<2c_i-1$ because $m<2^{n(m)}\le c_i$. Let $x=c_i+m$. The branch chain contains the construction of $\{x, 2x,4x,\dots\}$, which can obtain $tx$ based on BM. That is, according to the addition chain BM$(t)=\{a_0, a_1, a_2, \dots, t\}$, we obtain BM$_x(tx)=\{a_0x, a_1x, a_2x, \dots, tx\}$. The main chain and the root chain contain $\{1, 2, 4, \dots,c_i\}$, which can construct any integer from 1 to $2c_i-1$ by BM, including $C$ by BM($C$). As a result, using the ``backward-adding'' , each branch chain can generate an addition chain of \textit{e}. 
\par
\textbf{Theorem 2:} Let $l_{\mathit{SPTM}}$ be the length of addition chain obtained by SPTM. The range of $l_{\mathit{SPTM}}$ is
\begin{equation}
n(e)+2\sqrt{h(e)}-3 \le l_{\mathit{SPTM}} \le n(e)+h(e)-2.
\end{equation}
\par
\textbf{Proof: }In the worst case, all the branch chains can not get a shorter chain than the main chain. The method degenerates to BM, and the length is $n(e)+h(e)-2$.
\par
In the best case, all 1s in the binary form of $e$ are divided into identical form of $2^i+m$ (denote by $w$), and $h(e)$ can be factorized into $h(w) \cdot \frac{h(e)} {h(w)}$. For the first window \textit{w}, the length is $n(w)+h(w)-2$. For the other  $\frac{h(e)} {h(w)} -1$ windows, do $\frac{h(e)} {h(w)} -1$ times star step and $n(e)-n(w)$ times doubling step. Thus the addition chain length is $ n(w)+h(w)-2+\frac{h(e)} {h(w)} -1+(n(e)-n(w))=n(e)+h(w)+\frac{h(e)} {h(w)} -3 \ge n(e)+2\sqrt{h(e)}-3$, equality holds if and only if $h(w)=\sqrt{h(e)}$. 

\subsection{Cross Window Method }
\begin{figure}[htb]
\begin{floatrow}
\ffigbox{\includegraphics[width=0.3\textwidth]{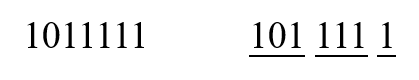}}
{\caption{Example of adjacent correlation.}
        \label{fig3}}

\ffigbox{\includegraphics[width=0.4\textwidth]{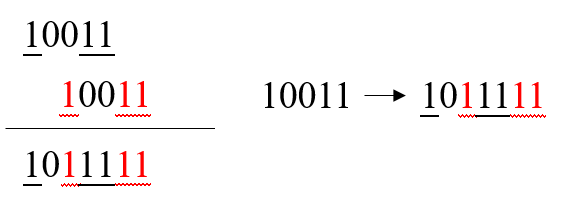}}
{\caption{Example of cross correlation.}
        \label{fig4}}
\end{floatrow}
\end{figure}

The Window method (WM) only considers the adjacent correlation which means that each window is divided sequentially. In practice, there are windows with cross correlation which means there is a cross relationship between the windows. Using cross windows can achieve better result in somes cases. For example, for the integer $(1011111)_2$, it only needs 2 star steps using cross windows, which is less than the result using adjacent windows, as shown in Fig. \ref{fig3} and Fig. \ref{fig4}.
\par
\par

Cross Window method (CWM) is to deal with the cross windows. CWM has two parameters:  valid window length $k$ and interval expansion length $s$. CWM has two parts, same as WM: pre-computation and construction. In pre-computation, the valid length $k$ is divided into two parts: the length of the right part $R= \lceil k/2 \rceil$, and the length of the left part $L=k-R \le R$.

\par
When $s \ge 1$, the pre-computation of CWM are constructed by inserting the interval expansion numbers between the left and right parts of the pre-computation of WM. The general binary structure is $(a)_2||0^s||(b)_2, a\in \{1, 2, 3, \dots, 2^L-1\}, b\in \{1, 3, 5, 7, \dots, 2^R-1\}$, which is performed specifically as follows:\par
(1) Get all odd numbers from 1 to $2^R-1$, and 2.\par
(2) Get the interval expansion numbers, which are $2^R, 2^{R+1}, \dots, 2^{R+s}.$\par
(3) Combine all numbers from 1 to $2^L-1$ with the interval expansion and the numbers in step (1).\par

\begin{figure}[htb]
\centering
\ffigbox{\includegraphics[width=1.0\textwidth]{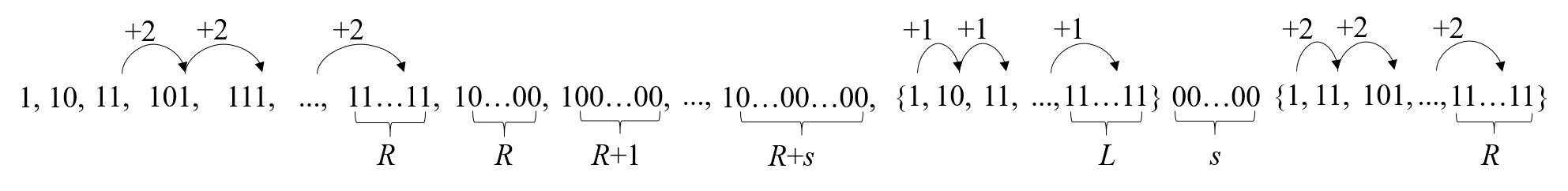}}
{\caption{The pre-computation of CWM by binary form.}
        \label{fig5}}
\end{figure}

Finally, the pre-computation ($s\ge 1$) is \{$1, 2, 3, \dots, 2^R-1; 2^R, 2^{R+1}, \dots, 2^{R+s}; 2^{R+s}+1, 2^{R+s}+3, \dots, 2^{R+s}+2^R-1; 2\cdot 2^{R+s}+1, 2\cdot 2^{R+s}+3, \dots, 2\cdot 2^{R+s}+2^R-1; \dots; (2^L-1)\cdot 2^{R+s}+1, (2^L-1)\cdot 2^{R+s}+3, \dots, (2^L-1)\cdot 2^{R+s}+2^R-1$\}, as shown by binary form in Fig. \ref{fig5}.\par
 The lengths in step (1),(2) and (3) are $2^{R-1}+1,s+1$ and $(2^L-1)2^{R-1}$. The total length is $2^{k-1}+s+2$.\par

When the interval expansion is not carried out (i.e. $s=0$), CWM degenerates to WM. In this case, step (2) should be removed. Thus the pre-computation ($s=0$) is \{$1, 2, 3, 5, 7, \dots, 2^k-1$\}, with length $2^{k-1}+1$, same as WM. It is unnecessary to divide the valid length $k$. For consistency, let $R=0$.
\par

\begin{figure}[htb]
\centering
\ffigbox{\includegraphics[width=0.35\textwidth]{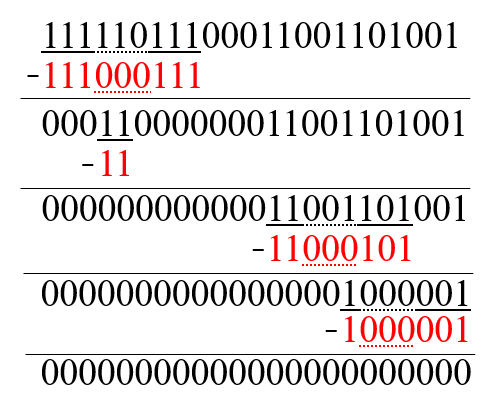}}
{\caption{An example of the process of CWM ($L=R=s=3$).}\label{fig6}}
\end{figure}

In the construction of CWM, for the binary form of \textit{e}, we read a window $w$ ($n(w)\le k+s$ and $MSB(w)=LSB(w)=1$). If $n(w)>s+R$, the interval expansion positions of the window are set to 0s. If $R<n(w)<s+R$, reset $w$ as its first \textit{R} bits with removing the tail 0s. If $n(w)<R$, do nothing. As a result,  \textit{w} is in the pre-computation. Then an operation $\ominus$ is performed. $y \ominus x$ is to align the highest non-zero bit of $y$ with the highest non-zero bit of $x$ and execute a subtraction.
 A concrete example is listed by binary form in Fig. \ref{fig6}.
\par
For $e=(11111011100011001101001)_2$, the first window is processed as $w_0=(111000111)_2$, then do $e=e\ominus w_0=(00011000000011001101001)_2$. Repeat this for $w_1=(11)_2,w_2=(11000101)_2$ and $w_3=(1000001)_2$. Finally, $e$ is zero.
From the above example, it is easy to find that $w_1$ is embedded in $w_0$ as $(111\underline{11}0111)_2$. As a result, it cannot construct the addition chain like WM. To solve this problem, we record all the windows at corrosponding locations and construct the addition chain from the recorded windows. That is, do doubling steps bit-by-bit from the first recorded window and add the window at each recorded position. The implementation of CWM is shown in Alg. \ref{alg5}. 

\begin{algorithm}[htb]
        \caption{Cross Window Method}
        \begin{algorithmic}[1] %每行显示行号
        \Require $k,s,e=(e_{n-1}\dots e_1e_0)_2$
        \Ensure $A = \{a_0, a_1, a_2, \dots, e\}$
        \If{$s \ge 1$}
        \State $A=$ \emph{pre-computation}$(s\ge1)$, $R=\lceil k/2 \rceil$
        \Else
        \State $A=$ \emph{pre-computation}$(s=0)$, $R=0$
        \EndIf
				\State $t=k+s,i=n-1,M = \{\underbrace{0,0,\dots,0}_n\}$
		\While{$i \ge 0$}   
				\If{$e_i=0$}
				\State $i=i-1$
				\Else
        \State find the longest bitstring $e_ie_{i-1}\dots e_j$ such that $i-j+1\le t$ and $e_j=1$
        \If{$i-j+1>s+R$}
				\State $w=(e_ie_{i-1}\dots e_{i-R+1}\underbrace{00\dots 00}_se_{i-R-s+1}e_{i-R-s}\dots e_j)_2, c=i-j+1-s-R$

         \EndIf		
        	\If{$R<i-j+1<s+R$}
				\State find the longest bitstring $e_ie_{i-1}\dots e_j$ such that $i-j+1\le R$ and $e_j=1$
				\State $w = (e_ie_{i-1}\dots e_j)_2, c=i-j+1$
         \EndIf
						
         \State $e = e\textcircled{-}w,M_{j}=w,i=i-c$

		\EndIf
    \EndWhile
    \State find the maximal $j$ such that $M_j\neq 0$
    \State $r=length(A),t = M_j,j=j-1$
		\While{$j\ge 0$}
		\State $A = A \cup \{ a_{r}:= 2t \}, r=r+1, t = a_{r-1}$
		\If{$M_j \neq 0$}
		\State $A = A \cup \{ a_{r}:= t + M_j \}, r=r+1, t = a_{r-1}$
        \EndIf
		\State $j=j-1$
        \EndWhile
        \State \Return $A$
        \end{algorithmic}
        \label{alg5}
    \end{algorithm}
\par

\textbf{Theorem 3:} Let $l_{\mathit{CWM}}$ be the length of addition chain obtained by CWM. The range of $l_{\mathit{CWM}}$ is
\begin{equation}
n(e)+2\sqrt{h(e)}-3 \le l_{\mathit{CWM}} \le n(e)+h(e)-2.
\end{equation}
\par
\textbf{Proof: }In the worst case, the position of 1s can not form any window with length longer than 1, which degenerates to BM, and the length is $n(e)+h(e)-2$.
\par
In the best case, like SPTM, all 1s are divided into several identical windows $w$, and the addition chain length is $n(e)+h(w)+\frac{h(e)}{h(w)}-3 \ge n(e)+2\sqrt{h(e)}-3$, equality holds if and only if $h(w)=\sqrt{h(e)}$. 
\par
\textbf{Theorem 4:} In general, let the number of recorded windows be \textit{v}, the length of pre-computation be $2^{k-1}+s+1+\beta$ ($\beta$ = 0 if $s=0$ otherwise $\beta=1$), and let the first window be $w_{0}$, the addition chain length obtained by CWM is
\begin{equation}
l_{\mathit{CWM}} = 2^{k-1}+s+\beta+n(e)-n(w_0)+v.	
\end{equation}
\textbf{Proof: } In CWM, we first construct the pre-computation. The length of pre-computation is $2^{k-1}+s+2$ if $s\ge 1$ otherwise is $2^{k-1}+1$, i.e. $2^{k-1}+s+1+\beta$ where $\beta$ = 0 if $s=0$ otherwise $\beta=1$. Then perform $n-n(w_0)$ times doubling step repeatedly and $v-1$ times star step for recorded windows except the first window. Thus the addition chain length obtained by CWM is
$
l_{\mathit{CWM}} =  2^{k-1}+s+1+\beta+n(e)-n(w_0)+v-1=2^{k-1}+s+\beta+n(e)-n(w_0)+v.	
$ 
\par

\subsection{Cross Window Method with Addition Sequence Algorithm}
The pre-computation of CWM can be optimized since some integers in the pre-computation may not be used as a window. In this paper, a new Addition Sequence algorithm (ASA) is presented to construct a short pre-computation of the used windows. ASA refers to the shortest addition chain containing given multiple integers, which is an \textbf{NP}-complete problem. However,  ASA is solvable in CWM, since only the pre-computation is involved which contains small integers. When we obtain a shorter pre-computation using ASA, we can also use larger valid window length and interval expansion length and are possible to obtain shorter addition chain. 
\par
Now we give a pragmatic ASA, which can find an addition chain containing all the used windows quickly. For an increasing order sequence $A =\{e_0,e_1,\dots,e_{d-1}\}$, let the last two numbers be $x,y (y>x)$ and let $y=tx+C (0\le C<x)$. For $tx$, we get BM$_x(tx)= \{a_0x, a_1x, a_2x, \dots, tx\}$ and put it in $A$ by increasing order. We put $C$ in $A$ by increasing order if it is not in $A$ and is non-zero. Thus the addition chain from $x$ to $y$ is formed. Repeat the above steps for the following two numbers in $A$ in reverse order until all integers in $A$ are solved. Finally, an addition chain containing $e_0,e_1,\dots,e_{d-1}$ is obtained. The implementation of ASA is shown in Alg. \ref{alg6}. 
\par
    \begin{algorithm}[htb]
        \caption{Addition Sequence algorithm}
        \begin{algorithmic}[1] %每行显示行号
        \Require $A =\{e_0,e_1,\dots,e_{d-1}\}$
        \Ensure \emph{an addition chain containing all integers in A}
        \State arrange $A$ in increasing order
        \State add 1, 2 into $A$ in increasing order if they are not in $A$
		  		\State $j=length(A)-1$
				\While{$j>1$}
		        \State $y=a_j,x=a_{j-1},y=tx+C (0\le C<x)$
				    \State add BM$_x(tx)$ into $A$ in increasing order
						\If{$C\neq 0$ and $C \not\in A$}
						  \State add $C$ into $A$ in increasing order
						  \State $j=j+1$
		         \EndIf
				    \State $j=j-1$
		    \EndWhile
		    \State \Return $A$
   		 \end{algorithmic}
   		 \label{alg6}
		\end{algorithm}

When the result of ASA is shorter than the original pre-computation in CWM (satisfied in most cases, but not absolutely), the original pre-computation will be replaced. CWM with ASA (CWM-ASA) is implemented in Alg. \ref{alg7}.
\par
\begin{algorithm}[htb]
        \caption{Cross Window method with Addition Sequence algorithm}
        \begin{algorithmic}[1] %每行显示行号
        \Require \textit{k,s,e}
        \Ensure $A = \{a_0, a_1, a_2, \dots, e\}$
        \State $A$ = CWM$(k,s,e)$, let all the windows used be a sequence $N$
        \State $\mathit{ASA\mbox{-}chain}$ = ASA$(N)$
        \If{$length(ASA\mbox{-}chain)< length(pre\mbox{-}computation)$}
  		\State replace the $pre\mbox{-}computation$ in $A$ with $\mathit{ASA\mbox{-}chain}$
		\EndIf
        \State \Return $A$
        \end{algorithmic}
        \label{alg7}
    \end{algorithm}
    
\textbf{Theorem 5:} 
Let $l_{\mathit{CWM\mbox{-}ASA}}$ be the length of addition chain obtained by CWM-ASA, the range of $l_{\mathit{CWM\mbox{-}ASA}}$ is the same as $l_{\mathit{CWM}}$, which is
\begin{equation}
n(e)+2\sqrt{h(e)}-3 \le l_{\mathit{CWM\mbox{-}ASA}} \le n(e)+h(e)-2.
\end{equation}
\par
\textbf{Proof: } 
The difference between CWM-ASA and CWM is the pre-computation. The lower bound of the pre-computation length in CWM-ASA is exactly the pre-computation length in CWM, thus the range of $l_{\mathit{CWM\mbox{-}ASA}}$ is the same as $l_{\mathit{CWM}}$. 
\par
\textbf{Theorem 6:} In general, let the number of recorded windows be \textit{v}, the length of pre-computation  be \textit{u}, and the first window be $w_{0}$, the addition chain length obtained by CWM-ASA is
\begin{equation}
l_{\mathit{CWM\mbox{-}ASA}} = u+n(e)-n(w_0)+v-1.	
\end{equation}
\par

\textbf{Proof: } In CWM-ASA, we first construct the pre-computation with length $u$ and then perform $n(e)-n(w_0)$ times doubling step repeatedly and $(v-1)$ times star step for recorded windows except the first window. Thus the addition chain length obtained by CWM-ASA is
$
l_{\mathit{CWM\mbox{-}ASA}} = u+n(e)-n(w_0)+v-1.	
$

\section{Numerical Results\label{chap:4}}
In this section, we implement BM, WM, SPTM, CWM and CWM-ASA and the performance are compared. We firstly show the performance of these methods on small integers with $l \le 22$. Then a general case is conducted with the integers generated randomly with different Hamming weight. Moreover, the integers of effective types of SPTM are exhibited to indicate the irreplaceable advantages of SPTM  in some cases. The parameters are selected as: WM: $1 \le k \le 20$; SPTM: $1 \le m \le 63$ and $m$ is odd; CWM: $1 \le k \le 10, 0 \le s \le 10$; CWM-ASA: $1 \le k \le 20, 0 \le s \le 20$. The final result for an integer of a method is the shortest addition chain length within the parameter range.
\subsection{The Integers with $l \le 22$}
For 365634 positive integers with $l \le 22$ \cite{TP-toolbox-web}, the results of BM, WM, SPTM, CWM and CWM-ASA are shown in Table \ref{table1}.

\par
%\begin{table}[htb]
%\linespread{1.2}\selectfont
%\setlength{\tabcolsep}{0.5mm}
%\begin{tabular}{|c|cc|cc|cc|cc|cc|}

%\caption{Results of all integers with $l \le 22$.}
%\label{table1}
%\end{table}
% \usepackage{multirow}
\begin{table}[H]
\linespread{1.2}\selectfont
\setlength{\tabcolsep}{0.5mm}
\begin{tabular}{|c|cc|cc|cc|cc|cc|}
% \usepackage{multirow}
\hline
\multirow{2}{*}{\begin{tabular}[c]{@{}c@{}}Gap with the\\ shortest\end{tabular}} & \multicolumn{2}{c|}{BM}             & \multicolumn{2}{c|}{WM}              & \multicolumn{2}{c|}{SPTM}            & \multicolumn{2}{c|}{CWM}             & \multicolumn{2}{c|}{CWM-ASA}          \\ \cline{2-11} 
                                                                                 & \multicolumn{1}{c|}{count} & prop.  & \multicolumn{1}{c|}{count}  & prop.  & \multicolumn{1}{c|}{count}  & prop.  & \multicolumn{1}{c|}{count}  & prop.  & \multicolumn{1}{c|}{count}  & prop.  \\ \hline
0                                                                                & \multicolumn{1}{c|}{7880}  & 0.0216 & \multicolumn{1}{c|}{46193}  & 0.1263 & \multicolumn{1}{c|}{68586}  & 0.1876 & \multicolumn{1}{c|}{86565}  & 0.2368 & \multicolumn{1}{c|}{228805} & \textbf{0.6258} \\ \hline
1                                                                                & \multicolumn{1}{c|}{31480} & 0.0861 & \multicolumn{1}{c|}{150463} & 0.4115 & \multicolumn{1}{c|}{187621} & 0.5131 & \multicolumn{1}{c|}{185261} & 0.5067 & \multicolumn{1}{c|}{131440} & 0.3595 \\ \hline
2                                                                                & \multicolumn{1}{c|}{66045} & 0.1806 & \multicolumn{1}{c|}{127654} & 0.3491 & \multicolumn{1}{c|}{100286} & 0.2743 & \multicolumn{1}{c|}{83943}  & 0.2296 & \multicolumn{1}{c|}{5379}   & 0.0147 \\ \hline
3                                                                                & \multicolumn{1}{c|}{81569} & 0.2231 & \multicolumn{1}{c|}{37075}  & 0.1014 & \multicolumn{1}{c|}{9051}   & 0.0248 & \multicolumn{1}{c|}{9597}   & 0.0262 & \multicolumn{1}{c|}{10}     & 2.7E-5 \\ \hline
4                                                                                & \multicolumn{1}{c|}{74605} & 0.2040 & \multicolumn{1}{c|}{4231}   & 0.0116 & \multicolumn{1}{c|}{90}     & 0.0002 & \multicolumn{1}{c|}{267}    & 0.0007 & \multicolumn{1}{c|}{0}      & 0.0000 \\ \hline
5                                                                                & \multicolumn{1}{c|}{59299} & 0.1622 & \multicolumn{1}{c|}{18}     & 4.9E-5 & \multicolumn{1}{c|}{0}      & 0.0000 & \multicolumn{1}{c|}{1}      & 3.0E-6 & \multicolumn{1}{c|}{0}      & 0.0000 \\ \hline
6                                                                                & \multicolumn{1}{c|}{26668} & 0.0729 & \multicolumn{1}{c|}{0}      & 0.0000 & \multicolumn{1}{c|}{0}      & 0.0000 & \multicolumn{1}{c|}{0}      & 0.0000 & \multicolumn{1}{c|}{0}      & 0.0000 \\ \hline
7                                                                                & \multicolumn{1}{c|}{13672} & 0.0374 & \multicolumn{1}{c|}{0}      & 0.0000 & \multicolumn{1}{c|}{0}      & 0.0000 & \multicolumn{1}{c|}{0}      & 0.0000 & \multicolumn{1}{c|}{0}      & 0.0000 \\ \hline
8                                                                                & \multicolumn{1}{c|}{3334}  & 0.0091 & \multicolumn{1}{c|}{0}      & 0.0000 & \multicolumn{1}{c|}{0}      & 0.0000 & \multicolumn{1}{c|}{0}      & 0.0000 & \multicolumn{1}{c|}{0}      & 0.0000 \\ \hline
9                                                                                & \multicolumn{1}{c|}{893}   & 0.0024 & \multicolumn{1}{c|}{0}      & 0.0000 & \multicolumn{1}{c|}{0}      & 0.0000 & \multicolumn{1}{c|}{0}      & 0.0000 & \multicolumn{1}{c|}{0}      & 0.0000 \\ \hline
10                                                                               & \multicolumn{1}{c|}{136}   & 0.0004 & \multicolumn{1}{c|}{0}      & 0.0000 & \multicolumn{1}{c|}{0}      & 0.0000 & \multicolumn{1}{c|}{0}      & 0.0000 & \multicolumn{1}{c|}{0}      & 0.0000 \\ \hline
11                                                                               & \multicolumn{1}{c|}{52}    & 0.0001 & \multicolumn{1}{c|}{0}      & 0.0000 & \multicolumn{1}{c|}{0}      & 0.0000 & \multicolumn{1}{c|}{0}      & 0.0000 & \multicolumn{1}{c|}{0}      & 0.0000 \\ \hline
12                                                                               & \multicolumn{1}{c|}{1}     & 3.0E-6 & \multicolumn{1}{c|}{0}      & 0.0000 & \multicolumn{1}{c|}{0}      & 0.0000 & \multicolumn{1}{c|}{0}      & 0.0000 & \multicolumn{1}{c|}{0}      & 0.0000 \\ \hline
AVG                                                                              & \multicolumn{2}{c|}{3.5433}         & \multicolumn{2}{c|}{1.4605}          & \multicolumn{2}{c|}{1.1369}          & \multicolumn{2}{c|}{1.0475}          & \multicolumn{2}{c|}{\textbf{0.3890}}          \\ \hline

\end{tabular}
\caption{Results of all integers with $l \le 22$.}
\label{table1}
\end{table}

In this range, from the first row, we can see the optimal results proportions of BM, WM, SPTM, CWM and CWM-ASA are 2.16\%, 12.63\%, 18.76\%, 23.68\% and 62.58\% respectively, and from the last row the average gap with the shortest is 3.5433, 1.4605, 1.1369, 1.0475, and 0.3890 respectively. The results of SPTM, CWM and CWM-ASA are better than those of BM and WM, and are more concentrated on the part with smaller gap. CWM-ASA has the best results, and the optimal and suboptimal (the gap with the shortest is 1) results account for 98.53\%.
\par

\subsection{The Integers Generated Randomly with Different Hamming Weight}
Let $p=\frac{Hamming\,weight}{bit\mbox{-}length}$, which means the bit 1 occurs with the probability of $p$. Select bit-length as 160, 384, 512, 1024, 2048, 4096 and $p$ as 0.1, 0.2, 0.4, 0.5, 0.6, 0.8, 0.9. Set 50 integers for each combination. The average addition chain lengths are shown in Table \ref{table2} by bit-length.

\par

\begin{figure}[htb]
\centering
\ffigbox{\includegraphics[width=0.8\textwidth]{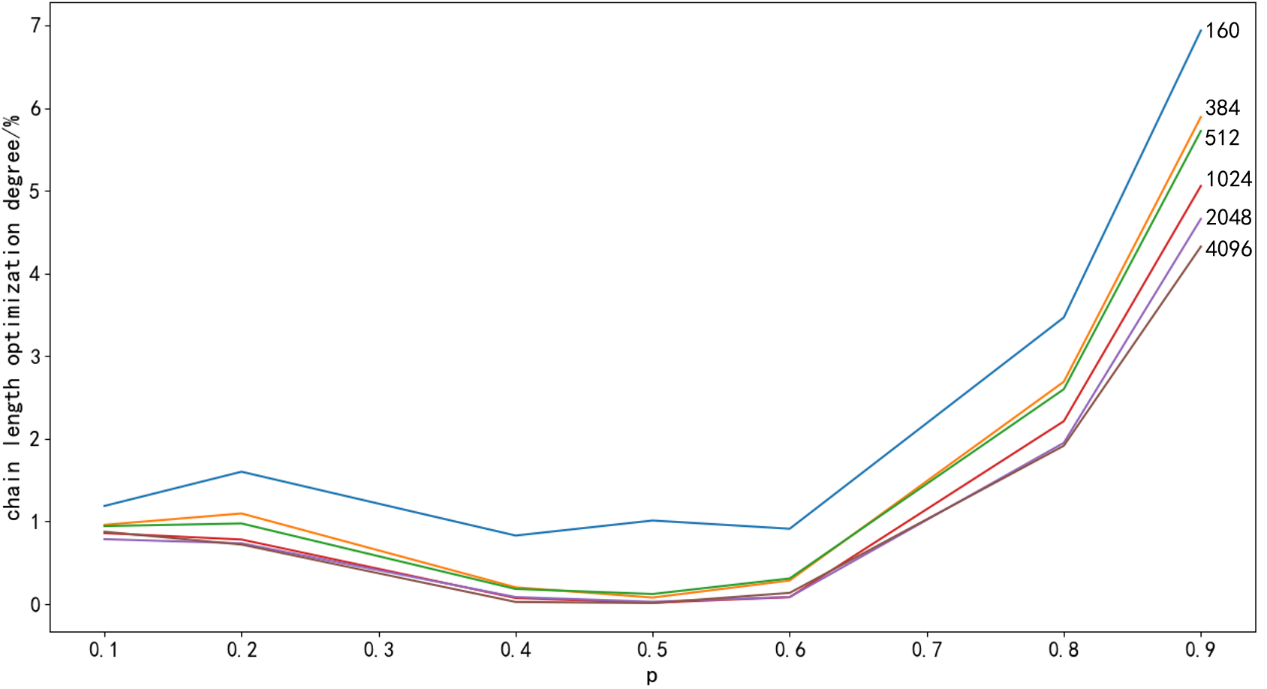}}
{\caption{Chain length optimization degree of CWM-ASA compared with WM.}
\label{fig7}}

\end{figure}

\begin{table}[htb]
\linespread{1.2}\selectfont
\begin{tabular}{|c|c|c|c|c|c|}
\hline
Len/bit & BM      & WM      & SPTM    & CWM     & \textbf{CWM-ASA}  \\ \hline
160  & 238.15  & 192.37  & 210.15  & 191.93  & \textbf{187.89}  \\ \hline
384  & 574.87  & 452.64  & 513.37  & 452.14  & \textbf{445.27}  \\ \hline
512  & 766.47  & 600.18  & 688.10  & 599.64  & \textbf{590.71}  \\ \hline
1024 & 1534.46 & 1182.50 & 1388.86 & 1181.94 & \textbf{1166.94} \\ \hline
2048 & 3070.68 & 2335.35 & 2793.60 & 2334.88 & \textbf{2307.24} \\ \hline
4096 & 6139.43 & 4620.42 & 5610.92 & 4619.92 & \textbf{4567.07} \\ \hline
\end{tabular}
\caption{Average addition chain lengths of random integers by bit-length.}
\label{table2}
\end{table}
In the random case, for each bit-length, the results obtained by SPTM is greater than WM, which shows that SPTM is not effective in this case. The results obtained by CWM and CWM-ASA are better compared to WM, and the chain length obtained by CWM-ASA is relatively short. Fig. \ref{fig7} shows the chain length optimization degree of CWM-ASA compared with WM.
\par
When $p\le0.5$, the optimization degree declines with the increasing of $p$, and the overall optimization degree is relatively low; when $p = 0.5$, the optimization degree is approximately the lowest; when $p\ge0.5$, the optimization degree generally increases with the increasing of $p$, and the overall optimization degree is relatively high.
\par
For the bit-length, with the increasing of the bit-length, the optimization degree of CWM-ASA declines. This is because the corresponding extra times doubling step are unavoidably brought in with the increasing of the bit-length, so that the overall cardinality becomes larger.
\par
In addition, the numbers with larger Hamming weight ($p = 0.95$) are tested, as shown in Table \ref{table3}.
For $p = 0.95$, the average optimization degree of CWM-ASA is 43.11\% compared with BM. When the bit-length is 4096, the average addition chain length obtained by BM is 7988.70, while CWM-ASA is 4414.82, and the optimization degree reaches 44.74\%.  The average optimization degree of CWM-ASA is 7.89\% compared with WM. When the length is 160 bit, the average addition chain length obtained by WM is 202.06, while CWM-ASA is 182.84, and the optimization degree reaches 9.51\%.

\begin{table}[htb]

\linespread{1.2}\selectfont
\setlength{\tabcolsep}{1mm}
\begin{tabular}{|c|c|c|c|c|c|}
\hline
Len/bit & BM      & WM      & CWM-ASA  & \begin{tabular}[c]{@{}c@{}}optimization\\ (CWM-ASA \& BM)\end{tabular} & \begin{tabular}[c]{@{}c@{}}optimization\\ (CWM-ASA \& WM)\end{tabular} \\ \hline
160     & 310.32  & 202.06  & 182.84  & 41.08\%       & 9.51\%        \\ \hline
384     & 746.76  & 470.42  & 430.36  & 42.37\%       & 8.52\%        \\ \hline
512     & 997.14  & 622.04  & 569.98  & 42.84\%       & 8.37\%        \\ \hline
1024    & 1994.92 & 1218.78 & 1126.40 & 43.54\%       & 7.58\%        \\ \hline
2048    & 3989.96 & 2394.72 & 2230.84 & 44.09\%       & 6.84\%        \\ \hline
4096    & 7988.70 & 4724.02 & 4414.82 & 44.74\%       & 6.55\%        \\ \hline
\end{tabular}
\caption{Comparison of CWM-ASA and BM, WM when $p = 0.95$.}
\label{table3}
\end{table}
\subsection{The Integers of Effective Types for SPTM}

SPTM is effective to the integers which have windows whose higest bit are followed by a long series of 0s and the rest of the windows. In this case, the length of the window is so long that the pre-computations of WM and CWM are overwhelming. Without using pre-computation, the result of SPTM is better.
\par
The generation rules of test integers are as follows:
\par
(1) Randomly select $4 \le \emph{k} \le 6$ and $40 \le \emph{s} \le 60$.
\par
(2) The integers of $k$ bits is generated randomly, and then one bit 1 and $s$ 0s are set ahead to form a window, and the window is copied to \emph{k} + 1 copies.
\par
(3) The positions of these windows are randomly generated, and the position distance among the windows is not less than \emph{k}. Then an integer is obtained from these windows with removing the tail 0s.
\par
The average test results are shown in Table \ref{table4}.
\par
\begin{table}[htb]
\linespread{1.2}\selectfont
\begin{tabular}{|c|c|c|c|c|c|}
\hline
Len/bit & BM      & WM      & SPTM             & CWM     & CWM-AS  \\ \hline
160     & 167.1   & 166.24  & \textbf{163.28}  & 165.48  & 165.34  \\ \hline
384     & 395.52  & 393.34  & \textbf{388.74}  & 392.6   & 392.42  \\ \hline
512     & 523.38  & 521.3   & \textbf{516.7}   & 520.6   & 520.44  \\ \hline
1024    & 1035.12 & 1033.08 & \textbf{1028.64} & 1032.56 & 1032.44 \\ \hline
2048    & 2059.1  & 2057.38 & \textbf{2052.62} & 2056.62 & 2056.54 \\ \hline
4096    & 4107.9  & 4105.26 & \textbf{4100.78} & 4104.78 & 4104.76 \\ \hline
\end{tabular}
\caption{Average test results of effective types for SPTM.}
\label{table4}
\end{table}
\par
In this case, the results obtained by SPTM are the best, and the results obtained by CWM and CWM-ASA are also better than those obtained by WM. This shows that although SPTM is not suitable for the random integers, it can achieve the best results among several methods for the windows having the higest bit followed by a considerable number of 0s.
\par
\section{Conclusion\label{chap:5}}
In this paper, we proposed a Simplified Power-tree method and a Cross Window method with a new Addition Sequence algorithm. The Simplified Power-tree method constructs a power-tree with deep deletion, which is more suitable when the windows have the highest bit followed by a considerable number of 0s. The Cross Window method considers the windows with cross relationship. The cross windows are processed by recording the window positions for recovery. Furthermore, the pre-computation is optimized with the Addition Sequence algorithm. The Cross Window method is slightly better than the Window method, and the Cross Window method with Addition Sequence algorithm has a better optimization, especially in the case of large Hamming weight. Roughly speaking, the average optimization degree is 7-8\%, and the best case is 9-10\%.

%\section*{Acknowledgment}
%This work was supported by Natural Science Foundation of Beijing %Municipality (No.
%4202037), NSF of China with contract (No.61972018).

%%%% 8. BILBIOGRAPHY %%%%
\bibliographystyle{alpha}
\bibliography{abbrev3,crypto,biblio}
%%%% NOTES
% - Download abbrev3.bib and crypto.bib from https://cryptobib.di.ens.fr/
% - Use bilbio.bib for additional references not in the cryptobib database.
%   If possible, take them from DBLP.

\end{document}